\documentclass[12pt,a4paper,aps]{article}
\usepackage{a4}
\usepackage{epsfig}
\usepackage[hmargin=2cm,vmargin=2cm,nohead]{geometry}
\begin{document}

%\begin{center}
%{\huge \bf 
\title{The Future of Nuclear Energy: Facts and Fiction \\ 
Chapter I: Nuclear Fission Energy Today}
%\end{center}

\author{
Michael Dittmar\thanks{e-mail:Michael.Dittmar@cern.ch},\\
Institute of Particle Physics,\\ 
ETH, 8093 Zurich, Switzerland\\
\date{August 3, 2009}
}
\maketitle
%\par
%\begin{center}
%{\large \bf 
%\author{Michael Dittmar \\ July,  2009}
%\end{center}

\begin{abstract}
Nuclear fission energy is considered to be somewhere between the holy grail, 
required to solve all energy worries of the human industrialized civilization, 
and a fast path directly to hell. Discussions about future energy sources and the possible contribution from 
nuclear energy are often dominated by variations of fundamentalists and often irrational approaches. 
As a result, very little is known by the general public and even by decision makers about 
the contribution of nuclear energy today, 
about uranium supplies, uranium resources and current and future technological challenges and limitations.

This analysis about nuclear energy and its contribution for tomorrow tries to shed light on the nuclear reality 
and its limitations in the near and long term future.  
The report, presented in four chapters,  is based essentially on the data provided 
in the documents from the IAEA (International Atomic Energy Administration) and the NEA (the Nuclear Energy Agency 
from the OECD countries, the WNA (World Nuclear Association) and the IEA (International Energy Agency).

\begin{itemize}
\item Chapter I summarizes the state of the world wide nuclear fission energy today and its perspectives for 
the next 10 years. 

\item Chapter II presents the situation concerning the secondary uranium and plutonium resources.  

\item Chapter III analyses the ``known" uranium resource data as presented within the past editions of the
IAEA/NEA Red Book.

\item In the final chapter IV, the plans and prospects for the long term future of nuclear fission and fusion 
will be analyzed. 
\end{itemize}
\end{abstract}

%\maketitle

\newpage
\section{Introduction}

Most people today think, without doubt, that a comfortable way of life depends on the availability of cheap energy with its 
almost limitless applications. The average per capita energy consumption in the  
developed world increased by a factor of three or more during the past 50 years. However, at most one billion people, 
or 1/7 of the human population of today, enjoy this increase.  
They live mainly in the richer countries and use on average roughly 50000 kWh thermal energy 
from various sources and per year. This is about 
three times higher than the world average consumption, about 5 times higher than the average per person energy use in  
China and 10 times larger than in India~\cite{IEAdatabasea}.

Depending slightly on the counting procedure, roughly 85\% of this energy comes from fossil energy sources: 
about 40\% from oil, 20\% from natural gas and 25\% from coal. Our mobility depends almost 100\%  on oil. Electric energy, made from 
various ``fuels" has the highest value for stationary applications and forms a basis for essentially all  hi-tech and luxury energy applications.
On a world wide scale, electric energy contributes to 16\% of the end energy use and between 20-25\% 
in most of the rich countries. About 70\% of the electric energy is again made from fossil fuels, about 16\% from 
hydropower and only 14\% from nuclear fission energy. The contribution from the renewable 
wind, solar and geothermal energy sources, despite some minor local exceptions, makes at best 1-2\%
\cite{IEAdatabaseb}. 

These numbers demonstrate that Electric energy, especially if made from nuclear energy and from renewable energy sources, 
contributes only very little to the real total world energy mix. 
In contrast to this reality, one obtains a totally distorted picture of their importance 
when one follows the media coverage and the political discussions at all levels  
about the pros and cons of nuclear fission power, hydropower, wind power, geothermal and direct and indirect solar energy sources.
For Switzerland, an interesting example for a small, densely populated and rich industrialized country,  
one finds that electric energy contributes to roughly 24\% of the final energy mix. 
It is made almost exclusively from hydropower ($\approx$ 60\%),  
and nuclear fission power ($\approx$ 40\%) \cite{Schweizexamp}.  Consequently the two big and three old
and smaller nuclear power plants contribute only 10\% to the swiss energy mix.

Similar basic energy numbers can be easily found on the internet from the IEA website~\cite{worldenergystat} and from  
many others. As fossil fuel resources, and especially oil and gas, are not renewable, it is obvious 
that the world wide energy mix of today is totally unsustainable.

While it is in general accepted that fossil fuels are not forever, the  
energy situation is mainly discussed in relation to the global warming effects.
This is reflected by various high level meetings, where climate change and other side effects 
of our energy use are on the agenda for world wide policy makers. 
Even though the recent price explosions for oil have resulted in some changes, 
the seriousness of very limited oil and gas resources are rarely discussed. If 
addressed one finds that they are discussed under the more ambiguous ``energy security" title.

Perhaps disillusioned by the official politically correct main stream arguments, many people have started 
to investigate the resource limitations, often under the title ``peak oil and gas" and ``peak everything".  
These problems and the need to react is now discussed at many levels and plenty of details 
can be found at web sites like the ``oil drum", the ``energy bulletin" and many others \cite{oildrumetc}.
Those who have accepted that the situation with our fossil fuel use is unsustainable suggest and support, in order  to prevent wars, 
chaos and collapse, mostly a mixture of the following three sometimes 
orthogonal evolutionary directions:
\begin{itemize}
\item The nuclear energy option;
\item The all renewable energy option, based dominantly on the transformation of solar and wind energy;
\item The energy reduction option, which stands for some efficiency improvements combined with an overall coordinated
reduction of consumerism. Consequently economic activities will slow down and ``we" all will have to live 
simpler, perhaps still satisfying, lifestyles.   
\end{itemize}

In this report we will shed some light on the nuclear energy option and its limitations.
This analysis is split into four chapters:  (I) the nuclear reality of today and its short term perspectives, (II) the situation concerning 
the secondary uranium resources, (III) the existing data about ``known" exploitable uranium resources 
and (IV) status and perspectives of fast breeder reactors (reactors of the fourth generation)
and why commercial fusion reactors will always be 50 years away.

We believe that a complete discussion 
should also address the problems related to (1) the real and imagined dangers of nuclear energy relative 
to other energy forms; (2) nuclear weapon proliferation and  
(3) the accumulated nuclear waste. Unfortunately it seems that political decision makers 
will not consider these issues as relevant for todays discussions about our energy future.
Therefore we will not enter into details here and refer the interested readers to the extensive literature 
about these issues\cite{bombwaste}. 
 
In this chapter nuclear energy and its place in today's world energy mix is reviewed. 
As significant new constructions in the nuclear power cycle, including uranium mines, 
enrichment facilities and power plants, require at least a 5-10 year construction time, the maximum possible 
contribution of the nuclear power sector up to almost 2020 is already known and presented 
in this report.

It should be clear from the facts presented in the following, 
that the nuclear energy situation is far from being in the often claimed nuclear renaissance phase.
In fact, even without the impact of the 2008/9 world financial chaos,  
it seems already very difficult to stop the slow nuclear phase out, with a annual decrease of about 1\%,  
observed during the past few years.  

\section{Energy from nuclear fission: \\
past, present and the next 10 years.} 

The physics laws which describe nuclear energy with its enormous energy density  
started being understood about 100 years ago, when a new form of very energetic 
radiation from heavy elements like uranium was discovered. It became quickly clear that 
a huge potential of applications were waiting to be discovered and used.
Among these applications, humans learned to build the ``final weapon of mass destruction" 
and found a way to produce commercial energy from the nuclear fission.  

In 1938 O. Hahn and F. Strassner studied the neutron bombardment of 
uranium observing some lighter elements. Within weeks,  L. Meitner and O. Frisch  could explain the 
reaction as the fission of uranium atoms into two lighter atoms. Today we know that on average 2-3 neutrons 
and a large amount of energy is liberated in this reaction.  
This observation opened the road to a controlled chain reaction using the neutrons 
emitted from one fission reaction to fission further uranium atoms. Such a chain reaction with a power of 2 Watt
was  first achieved by E. Fermi and his team in 1942. Only three years later the world saw  
the explosions of two fission bombs over Hiroshima and Nagasaki which killed immediately 150000 people.  

The civilian use of nuclear fission energy, or the nuclear epoch,  started during the 1950s
with the hope that this would lead mankind to an almost unlimited future energy supply. 
This idea came from the fact that the fission of 1 kg of the uranium isotope U235 liberates 
about the same amount of energy as one million kg of coal. 
Even if only the U235 component of natural uranium, which contains the two isotopes U238 (99.29\%) 
and U235 (0.71\%), can be used, one still finds that 1 kg of natural uranium 
contains the potential energy equivalent of 
more than 10000 kg of coal. Thus even a ``useless" rock, which contains perhaps only 0.01\% of uranium, 
e.g. 0.1 kg of uranium per ton, could in theory liberate more energy than 1 kg of coal. 

The necessary chain reaction to liberate nuclear fission energy is known to be possible if on average more than 
one neutron is emitted for every fission reaction. This is essentially possible only with 
two uranium isotopes, U235 and U233, and with the plutonium isotope Pu239 where on average 2-3 neutrons are 
emitted per fission reaction. Only U235 exists naturally in sizable quantities.

The fission of these heavy elements is induced usually by a bombardment with 
moderated (slowed down) neutrons. If these extra neutrons are used efficiently for new fission reactions, 
a chain reaction, either controlled (in a reactor) or uncontrolled (in a bomb) can be started.  
As only one neutron is required to keep the controlled chain reaction going, the other neutrons can 
be used to transform the non fissionable U238 and Thorium 232 isotopes under neutron absorption and 
subsequent decays into the fissionable isotopes Pu239 and U233.
This neutron absorption process can be used to breed (produce) fissionable material.
Already in the existing reactors, often called "once through" reactors, up to 1/3 of the produced power 
comes from the fission of Pu239 produced from the above U238 transformation.

A more complicated technological challenge is the Fast Breeder reactor. 
This reactor is operated in a chain reaction mode using the prompt energetic (fast) fission neutrons.
Theoretically the amount of fissionable material can be 
increased with fast breeders by a large factor. 
So far, prototype commercial Fast Breeder reactors 
were not a great success for energy production\cite{fastbreeder}. More details can be found 
in the Generation IV nuclear power plant road map\cite{gen4document}. 
In this document, written by scientists 
from all larger nuclear energy countries, it is stated 
that at least 20 years of intense research and development are required 
before the breeder option can be considered as a real alternative to 
the existing standard nuclear reactors.
More details about the status and prospects of Fast Breeders will be presented in chapter IV of this report.
   
\subsection{Nuclear fission power today}

Today, about 30 countries on our planet operate commercial nuclear fission power plants. 
During 2008, these power plants provided 2601 TWhe\cite{fission2008}\footnote{TWh stands for 
Terra Watt hours or $10^{12}$ Wh
and e for electric energy. The power of a standard nuclear power plant is usually given in the unit GWatt or $10^{9}$ Watt.
If a 1 GWatt reactor is operated with 85\% efficiency over one year, about 7.5 TWhe are produced.}. This is  
2.1\% less than in the record year 2006 where the world nuclear power plants produced 
2658 TWhe. As a consequence of the so far satisfied ever increasing electric energy demand,  
the contribution from nuclear fission energy to the total amount of produced electric energy 
has decreased from 18\% in 1993 to about 14\% in 2008. Roughly 16\% of the world energy end use
comes from electric energy\cite{IEAdatabaseb}. Multiplying this 14\% by 16\%,  
one finds that nuclear energy contributes now less than 2.5\% to the world's end energy mix.

This nuclear energy contribution is about a factor of three smaller than the one given in most reviews of the 
world energy situtions. For example the IEA and others 
try to combine various sources into the primary energy equivalent.
In order to do so, the produced thermal energy is used for the statistics and 
nuclear electric energy is multiplied by a factor of roughly 3.

However, this approach is somehow misleading as it is unclear on how hydropower,  where no thermal waste heat is produced,
should be used in comparison. Furthermore hydropower and gas fired power plants provide 
electric energy when needed. In contrast,  an efficient operation of 
nuclear power plants are requires their operation with little interruptions and at 100\% capacity.  
As a result, nuclear power plants produce the so called base load for the electric grid and hydro and gas fired power plants 
are used for the peak load needs. 

A fairer comparison would thus give the electric energy produced from hydropower a much higher quality factor
than the one from nuclear fission power.
Another problem with the primary energy accounting is related to the efficiency of nuclear power plants, 
which on average have a thermal to electric energy conversion factor of 33\%, much lower  
than modern fossil fuel power plants,  where efficiencies of  
50\% and more can be reached. In addition, 
the waste heat from nuclear power plants is of lower temperature 
than the one from gas power plants. Consequently, the usage of waste heat from todays nuclear power plants 
is much less efficient and therefore essentially wasted to the environment.
We thus find it more logical to measure the contribution from nuclear fission energy to the world end energy mix.

According to the IAEA database\cite{pris}, since the beginning of 2009, nuclear electric 
energy comes from 436 nuclear fission reactors with an electric power capacity of about 
370 GWe. The average age of these reactors is already about 25 years and 
roughly 130 reactors with a capacity of more than 90 GWe have an age between 30 and 40 years. 
A large fraction of those will probably be terminated during the coming 5 to 10 years.
The two oldest relatively small commercial reactors, 0.217 GWe each, 
have an age of 41 and 42 years and are expected to be shut down at the end of 2010\cite{WNAolduk}.

In contrast to the often repeated statement that the world is in a phase of a ``nuclear renaissance",
the data show a different picture.  Since the beginning of 2008, 
one reactor in Slovakia and two of the older reactors in Japan were terminated and not a single reactor 
was completed. In fact the year 2008 marks the first year, 
at least since 1968, when not a single new reactor was connected to the electric grid.
During the past 10-15 years, on average only about 3-5 nuclear power plants per year were connected 
to the electric grid and about the same number of smaller and older reactors were terminated. 

According to the IAEA data base\cite{pris}, currently 48 reactors are under construction and  
following the WNA data base perhaps 10 reactors per year will be completed during the coming 5-10 years\cite{WNAnew}.  
While the connection of about 10 reactors per year would indicate a substantial increase compared to the past 15 years, 
this number is far lower than 25 years ago when 33 reactors were connected.

If one assumes a usual reactor construction time of 5-10 years, one could imagine that all these 48 reactors  
might be operational between 2015 to 2020. If they can be operated as efficiently as the existing reactors,
these new nuclear power plants will contribute at most about 300 TWhe/year additional electric energy
resulting for the year 2015-2020 in total nuclear energy production of 2900 TWhe at most.

However, if one takes the average retirement age for the so far closed 122 reactors as a guideline, 
one can expect that up to 100 older smaller reactors will be terminated during the next 10 years. 
Combining these two pieces of information, it seems rather unlikely that  
even a net increase of the world wide fission produced number of TWhe is possible by 2015. 
In contrast, if one uses the decline of almost 1\%/year observed during the last years,  
a production of 2350 TWhe could be expected for the year 2015.

Consequently, one can predict for 2015, and ignoring other limiting factors, 
that the absolute contribution from nuclear power plants will remain, within $\pm$ 10\%, 
at the level of today.

Those interested to follow this nuclear evolution during the coming months and years 
can compare the planned and real start up dates which are summarized in a recent WNA 
reference document\cite{WNAnew}.
According to this WNA data base, it is planned that 7 and 8 new reactors will be connected to the 
grid during the remaing months of 2009 and 2010 respectively.
It seems that at least the 2009 expectations are already totally outdated.

\subsection{Requirements of natural uranium equivalent}

In the previous section we have presented how the long construction times for new nuclear power plants
and the existing age structure of nuclear power plants constrain the evolution of nuclear power during the next 5-10 years.
We will now investigate the nuclear fuel supply situation. 

Current nuclear reactors have, for several reasons, a relatively small thermal efficiency of about 33\%. 
To operate a 1 GWe power plant, one finds that U235 or Pu239 isotopes have to be fissioned 
at a rate of roughly $10^{20}$ fissions/sec (about 0.05 grams/second).
Knowing that the U235 isotope makes only 0.71\% of natural uranium, one finds that about 6.5 gram 
of natural uranium equivalent are required per second to operate a 1 GWe nuclear reactor.   
Multiplying this amount with the number of seconds per year one finds that 170 tons 
of natural uranium equivalent per year are needed to operate a 1 GWe power plant. Thus about 65000 tons/year of 
natural uranium equivalent per year are needed to 
operate the existing 370 GWe nuclear capacity. 
It is generally believed that (1) this amount of uranium can easily be 
obtained from the existing mines combined with secondary resources; 
(2) it will be easy to extract sufficient amount of uranium from new mines in the near future and (3)
no nuclear fuel shortages should be expected in the near future. 

However, as will be shown below and in chapter II of this review, the situation with uranium extraction from the known mines
and with the secondary resources during the coming 5 to 10 years appears to be much more critical than generally believed. 
Before we present these data,  a few more details about the usage of nuclear fuel might be helpful
to understand the current uranium supply situation and how it will constrain the evolution 
of nuclear power during the coming 5 to 10 years.  

Nuclear reactors produce energy from the fission of either 
uranium U235 or plutonium Pu239, which is one of the secondary sources of nuclear fuel.  
To simplify the discussion we always use the natural uranium equivalent in the following. 
As has been explained above, the amount of fissile material required to operate 
a 1 GWe nuclear power plant for one year, e.g.  assuming one annual refilling, is about 
165-180 tons of natural uranium equivalent per year. 
In practice, the normal operation of most reactors requires a few weeks of annual shutdown in order to replace
about 1/4 of the used up uranium fuel rods. 
Fresh reactor fuel rods contain a mixture of the fissile isotopes U235 or Pu239 component enriched to 3-4\% and 
U238. During the few years of operation the U235 content will be reduced to roughly 1\%. At the same time, 
due to neutron capture and subsequent $\beta$ decays some U238 is transformed into 
Pu239.  During the reactor operation Pu239 increases to something close to 1\% and contributes on average 
up to 30\% of the produced fission energy.  
Once the concentration of fissionable material in the fuel rods is reduced well below 2\% 
usually some new fuel rods are required. 
In case that one thinks about a large increase of new reactors in the future, it might be important 
to know that the first uranium load, which brings a new 1 GWe reactor to nominal power, 
is about 500 tons of natural uranium equivalent.

Some important statistics about nuclear power plants in different countries, their electric energy production in 2007 
and the corresponding uranium requirements, extracted  
from the Red Book data base of the IAEA and the NEA \cite{redbook07} and from the WNA, are summarized in Table 1.
The second column gives the number of reactors per country and in brackets the corresponding electric power.
The third column gives the total amount of electric energy produced in 2007.   
The number in brackets indicates the average number of TWh produced per installed GWe power 
which is an indication of how efficiently the nuclear 
power plants were operated in 2007\footnote{A non negligible number of reactors 
is always on some kind of long technical stop. A typical example is the result of the 2007 earthquake in Japan 
where some 8 GWe nuclear power plants were damaged and operation has not been resumed even two years later.}.
The number in the fourth column shows the natural uranium equivalent requirements for 2008. The number 
in brackets gives the average uranium requirements per GWe installed power for the world and the different countries. 
 
\small{
\begin{table}[h]
\vspace{0.3cm}
\begin{center}
\begin{tabular}{|c|c|c|c|}
\hline
Country         & 2007 number of                & 2007 produced                                       & 2008 uranium       \\
                     &  nuclear reactors       & electric energy  [TWhe]    & requirements [tons]          \\
                     &  (power [GWe])      &  (TWhe/per GWe power)   &    (per GWe [tons])       \\
\hline

World             & 439    (372)         &  2608 (7.0)                                &   64615    (174)          \\
USA               &  104    ( 99)       &    807   (8.2)                              &   18918       (191)        \\
France           &    59    ( 63)      &    420    (6.6)                             &   10527        (166)       \\
Japan            &     55   ( 48)     &    267      (5.6)                           &    7569          (159)      \\
Russia            &    31   ( 22)        &    148   (6.8)                              &     3365      (155)         \\
Korea (South) &     20  ( 18)  &     137        (7.8)                         &    3109           (177)     \\
Germany       &      17   ( 20)      &    133     (6.6)                             &    3332       (164)         \\
Canada         &      18   ( 13)     &        88    (7.0)                             &     1665      (132)          \\
Ukraine         &      15   ( 13)     &       87      (6.6)                           &    1974        (150)        \\
Sweden         &      10   (  9)    &      64        (7.1)                         &     1418         ( 157)    \\
China         &         18   (  9)     &      59        (6.9)                         &     1396         ( 163)     \\
UK                &       19   ( 11)     &      58      (5.2)                           &     2199        ( 199)       \\
Spain         &            8   (  7)    &      53        (7.1)                         &     1398          ( 188)    \\
Belgium       &          7   (  6)    &      46        (8.0)                         &     1011          ( 176)    \\
\hline
\end{tabular}\vspace{0.1cm}
\caption{Some basic data about the nuclear energy in the world and in some selected countries with more than 
1000 tons of annual uranium requirements\cite{WNA2007needs}.}
\end{center}
\end{table}
}
  
Since about 15 years only about 2/3 of the annual uranium requirements, 
between 31000 and 44000 tons are extracted from the world wide mining industry. This quantity is much smaller than 
the mining capacity, which for example in 2007 was, according to the Red Book, 54000-57000 tons\cite{RB07capacity}. 

The difference between the required and the extracted uranium in 2007 was about 23000 tons.
This is about the same amount as extracted by the three largest uranium producing countries, Canada, Australia and Kazakhstan. 
The missing amount of fissile material is currently satisfied with secondary resources. These are the civilian and military stocks of uranium and plutonium which were accumulated during the cold war and the so called MOX, 
a mixture of U235 and plutonium recycled in an expensive and technically challenging process 
from the used fuel rods. The tails left over from the U235 enrichment process  still contain 
some 0.2-0.3\% of U235 are another potential source of U235.   
In chapter II we will present the publicly available data on secondary resources, which provide some quantitative explanations  
for alarming situation which was expressed (highlighted by the author) 
in the IAEA and NEA press declaration from the 3rd of June 2008 about the new 2007 edition of the Red Book:

{\it  ``At the end of 2006, world uranium production (39 603 tonnes) provided about 60\% of world reactor requirements (66 500 tonnes) for the 435 commercial nuclear reactors in operation. The gap between production and requirements was made up by secondary sources drawn from government and commercial inventories (such as the dismantling of over 12 000 nuclear warheads and the re-enrichment of uranium tails). 
{\bf Most secondary resources are now in decline and the gap will increasingly need to be closed by new production. Given the long lead time typically required to bring new resources into production, uranium supply shortfalls could develop if production facilities are not implemented in a timely manner.}"}\cite{redbookpress}. 

\subsection{Uranium extraction, past and present}

In order to understand today's uranium supply situation, 
it is interesting to note that many formerly rich uranium mines, especially in 
large uranium consuming countries, are closed since many years.
This closure has happened despite that (1) the claimed goal is energy independence and
(2) uranium explorations make only minor contributions to the electricity price. 
Reality shows that these countries are now largely dependent on uranium imports from other countries.

Today, the ten largest uranium consumers are the United States, France, Japan, Russia, Germany, Korea (South), UK, Ukraine, 
Canada and Sweden. These countries require about 84\% of the world wide needed uranium or 
roughly 54000 tons of the natural uranium equivalent. This number can be compared with the uranium extracted world wide. 
The latest numbers from the WNA indicate that 43930 tons of uranium were extracted in 2008\cite{uran2008}
The corresponding data from the WNA and the Red Book 
for the previous years are 41279 tons in 2007, 39429 tons in 2006 and 41702 in 2005. 
Somewhat remarkable is the fact that the achieved numbers are usually  
at least one thousand tons smaller than the short term production forecast for the next year.

Only 4 of the above 10 countries, Canada, Russia, USA and Ukraine, are still extracting uranium in sizable quantities.  
Out of these four countries only Canada, which extracted 9476 tons in 2007, 
produces a large amount of uranium directly for export. It is interesting to note that the existing mines in Canada seem to be in 
a steep decline, while upgrades and new mines are unable to compensate this decline.  
During the years 2002-2005 the canadian mines produced on average more than 11000 tons per year.  
Since then, production fall by 5\% and more per year and only 9000 tons were produced in 2008.

The uranium mines in the above 10 largest uranium consumer countries, produce only about 28\% 
of their uranium needs or 15400 tons in 2007 and 14751 tons in 2008. If the two uranium exporting countries in this list 
are not taken into account, the remaining eight countries need to import about 95\% of their needs.
For the european countries the uranium import dependence is now almost 100\% and as such much larger 
than their relative dependence on oil and gas imports.  

As can be seen from Table 2, (East) Germany and France have essentially stopped uranium mining, 
even though they used to extract large amounts of uranium from within their territory. 
Finally, Japan, the UK, South-Korea and Sweden never had any substantial uranium mining on their territory. 

For the largest uranium consumer country, the United States, the situation is even more amazing.
The internal uranium production declined from a peak of up to 17000 tons per year around 1980 to a production of 
1654 tons in 2007 and 1430 tons in 2008. Last years amount does not even allow to operate 10\% of their nuclear power plants. 
More interesting questions should come up when one considers that currently 
about 50\% of the nuclear reactors in the USA are operated with 
the excessive military uranium stockpiles from Russia. 
As this contract ends in 2013 and as Russia has currently very ambitious plans to enlarge their own nuclear 
energy sector, it is most likely that the Russian authorities are already thinking about their own uranium needs.
Consequently, the stability of the electric grid in the United States became to some extend hostage 
to the friendship with their former archenemy and perhaps todays and tomorrows most important economic competitor.  
This USA dependence on Russia's ``good will" looks like an interesting problem for the next few years. 
These uranium data demonstrate the obvious contradiction between the  
goal that energy imports need to be reduced in order to achieve more energy security, as expressed by  
past and present US governments and reality. 

\small{
\begin{table}[h]
\vspace{0.3cm}
\begin{center}
\begin{tabular}{|c|c|c|c|c|}
\hline
Country         & nuclear                      & total uranium                   & produced/required       & peak production \\
                     &  electric power           & extracted                          &  2008   [tons]               & [tons] (year)        \\
                     &  2007 [GWe]               & up  to 2006 [tons]              &                                      &         \\
\hline
World             & 372                               &  2234083                                 &   43853/ 65000                   & 69692  (1980/81)     \\
USA               &  99             &    360401                                                    &     1430   / 18918                          & 16811  (1980/81)     \\
France           &  63.5           &    75978                                                     &   5 / 10527                              &  3394   (1987/88)    \\
Japan            &   47.6           &         84                            &   0/  7569                              &      10   (1972/73)    \\
Russia/FSU           &   21.7           &  132801                           &   3521 / 3365                              &      16000   (1987/88)    \\
Germany       &   20.3         &    219476                            &         0 / 3332                      &      7090   (1965/66)    \\
Korea (South) &  17.5          &       --                              &         0 / 3109                              &      --   (--)    \\
UK                &      11.0          &      --                                 &            0/ 2199               & -- (--)\\
Ukraine         &     13.1         &   12393                              &        800   / 1974             &   1000 (1992/93\\
Canada         &     12.6         &   408194                             &     9000 / 1665               & 12522 (2001/2002)\\
Sweden         &       9.0          &         200                                &            0/ 1418               & 29 (1969)\\
Australia                &      0         &    139392                                  &     8430/ 0               &   9512 (2004/05) \\
Kazakhstan               &      0          &   111755                                 &  8521/0               &  6654 (2007) \\
South-Afrika +  &   1.8          &  143194                                 &  5021/303           &  10188 (1980/81)  \\
Namibia  &                            &                                               &             &    \\
Niger                &          0          &      104087                                 &         3032/0               &   4363  (1981/82) \\

\hline
\end{tabular}\vspace{0.1cm}
\caption{Some important numbers about nuclear fission energy and present and past uranium production 
for the entire world and for different countries as given in the Red Book 2007\cite{redbook07} and the WNA data base\cite{redbookWNA}. }
\end{center}
\end{table}
}

{\bf Thus, the data demonstrate that there is nothing like uranium self sufficiency in the United States, the European Union,
Japan and other rich countries and that the uranium import dependence is in general much larger than for oil and gas.} 
In fact, the data on uranium mining and the large import dependence for several large uranium 
consuming countries undermines strongly the widespread believe 
that uranium resources are plentiful and that uranium exploration and mining costs are only a minor problem for 
nuclear energy production.  

A naive observer might conclude that the permanently repeated claims from  
authorities like from the NEA director general L. Ech\'avarri and the IAEA deputy director Y. Sokolov in 2006\cite{NEApress}, 

{\it ``Uranium Resources: Plenty to sustain growth of nuclear power"}

are either ``wishful thinking" or assume that such statements are required in order 
to keep the believe in a bright future for nuclear energy.

More details about uranium mining in different countries and especially their evolution during the past years and the near future needs 
will be presented in the next section.

\section{Uranium needs and production limits: the next 10 years} 
As we have seen in the previous section, for 2015 
the world nuclear power plants can reach a maximum capacity of 410 GWe.
In order to achieve this number it has to be assumed that none of 
current 370 GWe reactors will be terminated and that all plants, currently under construction,
can be completed by 2015. 

We will now estimate how much uranium fuel can be expected for the operation of nuclear power plants around the year 2015
and if this amount will provide a second constraint for the number of operating nuclear capacity.
Such estimates are rather reliable because the fuel needs for the reactor operating or under construction today 
are well known\footnote{Fuel requirements of future generation reactors are irrelevant for the next 10 years
as at least 20 years of research and development are required first \cite{gen4document}.}.

Nuclear capacity estimates and the corresponding uranium needs for the years beyond 2015 are becoming 
more and more speculative. For example one needs to know what will happen with the oldest 
nuclear reactors and if they can be replaced in time.
Nevertheless many government agencies, like the IAEA/NEA, the IEA or the EIA from the USA 
government and large pro nuclear organizations like the WNA
try to make forecasts at least up to the year 2030. 
For example the 2008 press declaration for the 2007 edition of the Red Book says \cite{redbookpress}:

{\it ``World nuclear energy capacity is expected to grow from 372 GWe in 2007 to between 509 GWe (+38\%) 
and 663 GWe (+80\%) by 2030. To fuel this expansion, annual uranium requirements are anticipated to rise to 
between 94 000 tons and 122 000 tons, based on the type of reactors in use today."} 

More generally, three scenarios for the evolution of the yearly nuclear capacity are 
envisaged for the next 20 years\cite{WNAgrowth}:
\begin{enumerate}
\item A fast growth with an increase of +2\% per year; 
\item the reference scenario with a 1\% growth per year 
and 
\item a slow decline scenario with a 1\% decrease per year which would start in 2010.  
\end{enumerate}

Taking the performance from the world wide nuclear power plants and from the uranium mines 
in the last few years as an indication only scenario (3), the slow phase out, seems to be consistent with the current data.
This trend might even be enforced by  the actual financial world crisis, which will make it even more difficult to obtain the  
large credits are needed for new nuclear power plants and new uranium mines 
and some construction delays for new nuclear projects have already been announced~\cite{usadelays}. 
In addition, one also knows that unpredictable events like earthquakes, accidents or wars can only 
result in a capacity decrease. 

In any case,  the upper limit uranium requirements and production capacities up to at least 2015 are already well known 
and summarized in Table 3 and 4. As quantified within the Red Book 2007 edition and the WNA 2009 data base, 
the expected increase in nuclear power plant capacity is expected to come from a few countries only.  
Some important aspects about these near future world wide nuclear plans are:  
\begin{itemize}
\item Germany, the fifth largest nuclear power country, has given a definite plan for their nuclear phase out. 
According to this plan by 2015 the german nuclear power capacity should be reduced from  
20.3 GWe to about 11 GWe~\cite{germany}.
\item Very ambitious plans~\cite{WNAnew} to complete a large number of nuclear power plants by 2015  
are currently proposed by China where the current 7.6 GWe (2007) should increase to 25-35 GWe. 
A similar increase is planned by India where 3.8 GWe (2007) should increase to 9.5-13.1 GWe.
This can be compared with the plans from 
Japan, Russia and South Korea, where their entire capacity should increase by an additional 8-10 GWe.
\item The rich OECD countries are planning currently for a roughly constant nuclear capacity.
\end{itemize} 

However, the high growth 2010 forecast from the IAEA/NEA Red Book 2007 is, according to the 
more recent March 2009 WNA numbers, already unachievable. 
In fact even this WNA estimate, which assumes that during 2009 and 2010
seven (4.3 GWe) and eight (5.2 GWe) new nuclear power plants 
will be connected to the grid\cite{WNAnew}, seems to be totally unrealistic. 

\small{
\begin{table}[h]
\vspace{0.3cm}
\begin{center}
\begin{tabular}{|c|c|c|c|}
\hline
source                     & power    [GWe] 2010                  &  power [GWe]  2015                      & power [GWe]  2009      \\
\hline
2007 edit.    world total            &  377-392              &   410-456                            &   370         \\
2007 edit.    OECD only           & 304-309               &    310-326                           &   304         \\
WNA March 2009 World          &  $\leq$ 380            &   440                                   &   370                      \\
\hline
%Country                   & power 2010           & power 2010 (2015) & 2007 +2010 (+2015)              \\
%                               &  [GWe] RB05 2005             & [GWe] RB07 2007       &  [GWE] RB07 +WNA09  \\
\hline
China                      & 13-20                                            & 25-35          &  7.6   \\
Germany                 & 12.5-14.5                                        & 8-12         &   20.3    \\
India                        &  6.2-6.7                                        &  9.2-13.1      &  3.8   \\
Japan                      & 48.5                                          & 49.8-55.0        & 47.1  \\
Korea (S)                 & 17.5-18.2                                  & 24.1-25.5       & 17.4   \\
Russia                     & 24-25                                           & 30-32      & 21.7        \\
other                       &  256-257                                   & 264-284  &  262           \\               
\hline
\end{tabular}\vspace{0.1cm}
\caption{Nuclear power perspectives up to 2015 for different countries and the world as given 
by the Red Book 2007\cite{redbook07}. The WNA numbers are taken from \cite{WNAnew}.}
\end{center}
\end{table}

As we are interested in estimating the maximum possible contribution from nuclear power plants 
during the next decade, the above Red Book scenario can be used as a guideline to estimate the 
requirements of uranium equivalent for the coming years. In order to operate the 
current and future running nuclear reactors, the authors of the Red Book 2007 estimated that 
between 70-75 000 tons of uranium equivalent are required for the year 2010 and 
between 77-86000 tons by 2015. Following the IAEA/NEA June 2008 press declaration, already quoted
in section 2.2, such a growth for uranium mining seems to be a serious challenge:

{\it ``Given the long lead time typically required to bring new resources into production, uranium supply shortfalls could develop if production facilities are not implemented in a timely manner.}"\cite{redbookpress}. 

Despite this and similar more hidden warnings, the authors of the Red Book usually give a 
rather rosy picture for the future uranium mining
as can be seen from the data summarized in Table 4. The expected large increase of world wide uranium 
mining should almost exactly match the requirements. However, and as in past years, essentially all countries 
seem to exaggerate the mining capacities far beyond the amount 
which can be extracted as demonstrated from the comparison of 
the 2007 claimed capacity and the real achieved uranium 2007 mining results. 
The numbers in the first column indicate especially large differences 
for Canada and the USA.
 
For 2007 the world wide uranium mining capacity is given as 
of 54370-56855 tons. In comparison to this capacity , the expectations from the Red book for the year 2007 
were given as 43328 tons. Uranium mining 2007 achieved 41264 tons, about 2000 tons smaller than the 
forecast for the same year. 
Similar wrong estimates were given in past Red Book editions.
For example the Red Book 2003 (2005) gave capabilities for the year 2003 (2005) of 49940 tons (49720-51565 tons).
In comparison, the achieved uranium extraction was 35492 tons in 2003
and 41943 tons in 2005\cite{olderredbooks}.  

As if these capacity numbers, exaggerated by 20-30\%, would not be troubling enough, the discrepancy 
between the claimed new mining capacity and what is really achieved is even more surprising. 
According to the Red Book 2007, the total additional capacity in 2007 compared to 2005 was estimated as 5290 tons.
The real result for 2007, a combination from older sometimes declining operating mines 
and the new mines,  was about 700 tons lower than the one achieved in 2005. For 2008 the production reached 
43930 tons which is 2200 tones larger than in the year 2005 but still far below the increase expected already for 
during 2007\cite{uran2008}.

Similar discrepancies between Red Book predictions and what could be extracted can be found from past Red Book editions. 
These discrepancies are, somewhat hidden, acknowledged in the latest 2007 edition. 
Unfortunately, instead of explaining the origin of such mistakes and correcting them, 
in order to improve the quality of the Red Book, it seems that systematic differences are simply accepted with  
the statement that{\it {\bf``World production has never exceeded 89\% of the reported production capability and since 2003 
has varied between 75\% and 84\% of production capability"}}\cite{Redbookerror}.
Further inconsistencies exist between the expected mining capacity increase and the detailed 
timetable given for the opening and extensions of uranium mines \cite{Redbookcap}.
For example the Red Book forecast, Table 24 (page 48), assumes that between 2007 and 2010, 
the uranium mining capacity will increase by 26000 to 29900 tons. 
However, a direct counting of the new uranium mines (page 49) results in a new capacity of about 20000 tons.

Similarly the forecast between 2010 and 2015 assumes that new mining projects should increase the capacity 
by another 15-30000 tons. In comparison, the direct counting of new uranium mines sums up at best to about 21000 tons,  
about 30\% below the claimed upper value of 30000 tons. 

A critical reader of the Red Book will thus be intrigued to investigate in which countries these capacity increases 
are expected. Some of these predictions, extracted from the Red Book, are shown in Table 4 below. 
One finds that about 50\% of the world wide uranium increase between 2007 and 2010  
should come from Kazakhstan. It is claimed that their production capacity will increase from 7000 tons in 2007 
to 18000 tons. Such an increase should have raised some critical reflections and comments from the authors of the Red Book  
as it would put Kazakhstan on equal terms with the combined production from Canada and 
Australia in 2008. According to the WNA spring 2009 document about Kazakhstan the 2010 forecast has  
already reduced to 15000 tons~\cite{kazakhstan}. If one takes the latest news about a huge corruption affair concerning the 
uranium resources of Kazakhstan into account~\cite{wnanews}, a drastic reduction of the 2009 and 2010 forecasts 
can be predicted.

Uranium mining in Canada also seems to be far behind the Red Book expectations\cite{canada}.
Not only are the real mining numbers much lower than the claimed capacities, 
but the existing three mines, which produce essentially 100\% of the uranium, seem to be in steep decline.
The production from these three large mines (McArthur River, McClean Lake and Rabbit Lake)
declined from 11400 tons in 2005 to 9000 tons in 2008. 
The previously expected 2007 start of the Cigar Lake mine, with an estimated yearly production capacity of 7000 tons, 
was stopped due to a catastrophic flooding in late 2006. The startup date of this 
mine is now delayed at least until 2012.

One might conclude that the Red Book uranium mining extrapolations are exaggerated and not based 
on hard facts as one would have expected from this internationally well respected document.
    
\small{
\begin{table}[h]
\vspace{0.3cm}
\begin{center}
\begin{tabular}{|c|c|c|c|}
\hline
Red Book                     & 2007 mining ratio                       &  prod. capacity                     & prod. capacity        \\
2007                             & production/capacity         & 2010 [1000 tons]                 &  2015  [1000 tons]    \\
\hline
World                                       & 0.73-0.76              &  80.7-86.7             &   95.6-117.4                                        \\ 
\hline
world forecast                    & ---    &  new capacity 2007-2010 & new capacity 2010-2015                      \\
new mines                          & ---    &  26.3-29.9   [1000 tons]    &  15.0-30.7    [1000 tons]                           \\
\hline
Australia                       & 0.91                           &    10.2                      &    10.2-19.0                                                \\
Canada                        & 0.63                           &      17.7-19.3            &  17.7-19.3                                              \\
Kazakhstan                  &  0.94                                 &     18.0                    &  21.0-22.0                                                \\
Namibia                       & 0.58                          &        6.0-7.0               & 8.0-9.0                                                        \\
Niger                           & 0.8                            &         4.5                     & 10.0                                                     \\
Russia                         & 1.0                           &         4.7-5.0               & 7.4-12.0                                                 \\
USA                            & 0.37-0.59                  &          3.4-6.1               & 3.8-6.6                                                \\
\hline
Scenario A             &     ---                                 &  60.5-65.0             &   72-88.0                   \\
Scenario B             &     ---                                 &  53 - 55                               &    61.5-70                                              \\
\hline
\end{tabular}\vspace{0.1cm}
\caption{Expected uranium production capacity given in 1000 tons from the Red Book for world and for different countries and for 
the years 2010 and 2015\cite{Redbookcap}. The expected world wide capacity increase between 2007 
and 2010 and from 2010 to 2015 are obtained from the evolution of the total capacity.
The ratio between the real production numbers for 2007 from the WNA and the uranium capacity from  
the Red Book are given in column 2. Scenario A and B are a rough forecasts
for the maximal uranium mining for the years 2010 and 2015 and based on the past capacity and real mining relation.
For Scenario A it assumed that the mining performance will be 75\% of the future capacity expected according to the Red Book.  
For Scenario B we assume that the existing mines in 2007 will continue an average annual production of 40000 tons and  
that only 50\% of the capacity forecast can become operational in time.}
\end{center}
\end{table}
}

Those who are interested in the near future nuclear energy contribution and thus 
uranium mining perspectives for the next 10 years should consequently not use the Red Book data directly.
Instead one could try to guess more realistic numbers by using the ratio between the 2007 mining results 
and the 2007 capacity as a first guess and update and improve these numbers accordingly during the next few years.  
Following this method, one would reduce the mining capacities by at least 20-30\% in order to obtain some forecast
(Scenario A). As a result one would predict a total uranium production of 
about 60000 tons in 2010 and 72000 tons in 2015. 
At least for 2010 Scenario A numbers are too high.

For Scenario B, we use the evolution of new uranium mines in order to 
determine how fast new capacity can become operational. 
Using this procedure and the real mining data from the past few years,  roughly 40000 tons per year,
and assuming that only 50\% of the new mining capacities can be realized 
one would predict a perhaps more production of 54000 tons in 2010, 61500 tons by 2015.
Those numbers can be compared to the latest WNA June 2009 estimates, where a total of 49400 tons 
and 74000 tons are predicted for 2009 and 2015 respectively~\cite{WNAfuture}. It seems that such professional estimations 
do not use much more input than a mixture of the above two simple minded methods. 
Within less than one year one will be able to update the above scenarios using the  
2009 results and improve the 2010 and 2015 forecasts\footnote{For those interested, I am offering a bet that 
the 2009 and 2010 numbers will not be higher than 45000 tons and 47000 tons respectively.}.

Taking into account that civilian secondary resources currently provide about 
21000 tons of natural uranium equivalent per year and that the civilian 
part of these resources will be basically exhausted within the next few years one finds 
that even the optimistic WNA 2009 numbers indicate uranium fuel supply stress during the coming years.
According to a recent presentation at the annual WNA September 2008 symposium 
from the Ux consulting (Macquarie Research commodities )~\cite{WNAfuture} 
about 1200 tons of uranium are missing for the 2009 demand.  
Furthermore, an uranium mining result 
below 50000 tons/year in 2009 and beyond will result in a serious uranium shortage. 

\section{Summary Part I:Nuclear fission energy today}

Our analysis of the publicly available data from the large international and very pro nuclear organizations, the IAEA 
and the WNA, show that the current evolution of nuclear fission energy is consistent with a slow nuclear phase out.
This situation is summarized by the following points: 

\begin{itemize}
\item The overall fraction of nuclear energy to electric energy has gone down from 18\% in 1993 
to less than 14\% in 2008. With electric energy providing roughly 16\% of the world wide 
energy end use one finds that overall the nuclear energy contribution of less than 2.5\%.
\item The number of produced TWhe electric energy from 
world wide nuclear power plants is now lower than in 2005 and it has decreased by about 2\%
from a maximum of 2658 TWhe in 2006 to 2601 TWhe in 2008. 
\item Today and world wide, 48 nuclear power plants with a capacity of about 40 GWe 
are under construction. Only 10\% of them are being constructed within the OECD countries, 
which host currently about 85\% of the existing nuclear reactors.
However, about 100 older reactors with slightly larger capacity 
are reaching their retirement age during the same period. It follows that 
even if all 48 reactors might be connected within the next 5 to 10 years to the electric grid, 
it will be difficult to keep the current level of TWhe produced by nuclear energy. 
\item The natural uranium equivalent required to operate the 370 GWe nuclear power plants of today
is roughly 65000 tons per year. However, during the past 10 years, the world wide uranium mining extracted on average only 
about 40000 tons of uranium per year and the difference had to be compensated 
by the secondary resources. According to the data from  the Red Book 2007 and the WNA
the remaining civilian uranium stocks are expected to be terminated during 
the next few years. Consequently the current uranium supply situation is unsustainable. 
\item The urgency to increase world wide uranium mining by a large number
is well documented in the current and past Red Book editions and the related official declarations.
However, the latest uranium mining data indicate that new uranium mines will not be capable to
compensate for the diminishing secondary uranium resources and that it will be difficult to  
fuel the existing 370 GWe. It seems that either a rather welcome but improbable further 
large conversion of nuclear weapons into reactor material will happen 
during the coming years, or fuel supply problems within the next 3-5 years will force a 10-20 GWe reduction of 
the operational nuclear power capacity.   
\end{itemize}

We can thus conclude this chapter I, {\bf Nuclear Fission Energy today}, 
with the statement that none of the publicly available official data support the 
widespread belief that the world is in a ``Nuclear Energy Renaissance" phase. 
In reality, the data about uranium mining and the large number of aging nuclear reactors 
indicate that the trend of a 1\% yearly decrease of the number of fission produced TWhe
will continue at least up to 2015.  In fact, the increasingly serious uranium supply situation 
might lead even to a forced nuclear shutdown of perhaps 5\% of the world wide reactors,  
most likely in countries without sufficient domestic uranium mining and enrichment facilities.
Such a result would certainly end the widespread belief in a bright 
future for nuclear fission energy.

%\vspace{1.cm}
%\noindent
%{\bf \large Acknowledgments} \\
%{\normalsize  \it{
%I would like to thank 
%}
%}

%\newpage

\end{document}